\documentclass[12pt]{article}
\begin{document}

\title{{\bf Born's Rule Is Insufficient \\in a Large Universe}
\thanks{Alberta-Thy-03-10, arXiv:yymm.nnnn [hep-th]}}

\author{
Don N. Page
\thanks{Internet address:
don@phys.ualberta.ca}
\\
Theoretical Physics Institute\\
Department of Physics, University of Alberta\\
Room 238 CEB, 11322 -- 89 Avenue\\
Edmonton, Alberta, Canada T6G 2G7
}

\date{2010 March 4}

\maketitle
\large
\begin{abstract}
\baselineskip 25 pt

Probabilities in quantum theory are traditionally given by Born's rule as the
expectation values of projection operators.  Here it is shown that Born's rule
is insufficient in universes so large that they contain identical multiple
copies of observers, because one does not have definite projection operators to
apply.  Possible replacements for Born's rule include using the expectation
value of various operators that are not projection operators, or using various
options for the average density matrix of a region with an observation.  The
question of what replacement to use is part of the measure problem in cosmology.

\end{abstract}

\normalsize

\baselineskip 23 pt

\newpage 

{\bf Probabilities in quantum theory are traditionally given by Born's rule
\cite{Born}.  This rule says that probabilities are the absolute squares of
quantum amplitudes.  More precisely, Born's rule gives probabilities of
measurement or observation results as the expectation values of a complete
orthogonal set of projection operators.  This rule seems to work well for
ordinary laboratory settings, where one is considering the observations of a
specific observer and knows where he or she is within the quantum state. 
However, the universe may be so large that it contains identical multiple copies
of the observer and the measurement situation, so that the observer does not
know which copy he or she is.  Here I show that Born's rule is insufficient in
such cases.  Normalized probabilities for the outcomes that can be distinguished
by a local observer cannot be given by the expectation values of any projection
operators in a global quantum state of the universe.  There are several possible
replacements for Born's rule, such as using the expectation values of various
operators that are not projection operators.  The measure problem in cosmology
\cite{BLL94,Vil95a,cmwvw} is a reflection of the uncertainty of what the correct
rule is.}

Traditional quantum theory uses Born's rule for the probability of an
observation $O_j$ (the result of an observation) as $P_j \equiv P(O_j) = \langle
\mathbf{P}_j \rangle$ where $\mathbf{P}_j$ is the projection operator onto the
observational result $O_j$, and where $\langle\ldots\rangle$ denotes the quantum
expectation value of whatever operator replaces the $\ldots$ inside the angular
brackets.  Born's rule works when one knows where the observer is within the
quantum state (e.g., in the quantum state of a single laboratory rather than of
the universe), so that one has definite orthonormal projection operators. 
However, Born's rule does not work in a universe large enough that there may be
identical copies of the observer at different locations, since then one does not
know uniquely where the observer is or what the projection operators are.

For example, suppose there are two copies of the observer, at locations $A$ and
$B$.  The two copies are assumed to be identical, by which I mean all local
observations the observer might make cannot distinguish them.  The two copies
may be distinguished globally by their different locations, but that information
is not available to the copies of the local observer themselves.  Suppose, for
simplicity, that each copy of the observer makes an observation that can give
either the result 1 or 2, with no other possibilities.  One would like a theory
of the universe (including a specification of its quantum state) that would give
normalized probabilities of getting the results 1 and 2, say $P_1$ and $P_2$
respectively, without having to specify the inaccessible information of what the
location is.

Born's rule would give the probabilities $P_1^A = \langle \mathbf{P}_1^A
\rangle$ and $P_2^A = \langle \mathbf{P}_2^A \rangle$ if the observer knew that
it were at location $A$ with the projection operators there being
$\mathbf{P}_1^A$ and $\mathbf{P}_2^A$.  Similarly, it would give the
probabilities $P_1^B = \langle \mathbf{P}_1^B \rangle$ and $P_2^B = \langle
\mathbf{P}_2^B \rangle$ if the observer knew that it were at location $B$.  (All
these projection operators act on the full quantum state, but they act
nontrivially only at their respective locations $A$ and $B$.  For simplicity we
shall assume that the two locations are at spacelike separations, so that the
two sets of projection operators commute with each other.)

However, if the observer is not certain to be at either $A$ or $B$, and if
$P_1^A \neq P_1^B$, then neither $P_1^A$ nor $P_1^B$ would be the probability
$P_1$ of simply getting the observational result 1.  I shall assume that $P_1$
must be a weighted mean of $P_1^A$ and $P_1^B$ with both weights positive, and
so be strictly between $P_1^A$ and $P_1^B$.  However, there is no
state-independent projection operator that gives an expectation value with this
property for all possible quantum states, as I shall now prove.  (If one were
allowed to choose the projection operator to depend on the quantum state, then
one could get any expectation value one wanted from any quantum state, so I
shall exclude that possibility.)

Consider normalized pure quantum states of the form
\begin{equation}
|\psi\rangle =  b_{12}|12\rangle + b_{21}|21\rangle, 
\label{2-state}
\end{equation}
with arbitrary normalized complex amplitudes $b_{12}$ and $b_{21}$.  The
component $|12\rangle$ represents the observation 1 in the region $A$ and the
observation 2 in the region $B$; the component $|21\rangle$ represents the
observation 2 in the region $A$ and 1 in the region $B$.  Therefore, $P_1^A =
P_2^B = |b_{12}|^2$, and $P_2^A = P_1^B = |b_{21}|^2$.

For Born's rule to give the possibility of both observational probabilities'
being nonzero in the two-dimensional quantum state space being considered, the
orthonormal projection operators should each be of rank one, of the form
\begin{eqnarray}
\mathbf{P}_1 = |\psi_{1}\rangle\langle\psi_{1}|,\ 
\mathbf{P}_2 = |\psi_{2}\rangle\langle\psi_{2}|,
\label{projections}
\end{eqnarray}
where $|\psi_{1}\rangle$ and $|\psi_{2}\rangle$ are two orthonormal pure states.

However, once the state-independent projection operators are fixed, then if the
quantum state is $|\psi\rangle = |\psi_{1}\rangle$, the expectation values of
the two projection operators are $\langle\mathbf{P}_1\rangle \equiv
\langle\psi|\mathbf{P}_1|\psi\rangle =
\langle\psi_1|\psi_{1}\rangle\langle\psi_{1}|\psi_1\rangle = 1$ and
$\langle\mathbf{P}_2\rangle \equiv \langle\psi|\mathbf{P}_2|\psi\rangle =
\langle\psi_1|\psi_{2}\rangle\langle\psi_{2}|\psi_1\rangle = 0$.  These extreme
values of 1 and 0 are not positively weighted means of $P_1^A$ and $P_1^B$ and
of $P_2^A$ and $P_2^B$ for any choice of $|\psi_{1}\rangle$ and
$|\psi_{2}\rangle$ and any normalized choice of positive weights.  Therefore, no
matter what the orthonormal projection operators $\mathbf{P}_1$ and
$\mathbf{P}_2$ are, there is at least one quantum state (and actually an open
set of states) that gives expectation values that are not positively weighted
means of the observational probabilities at the two locations.  Thus Born's rule
fails.  This proof is simpler and uses much weaker assumptions than my previous
arguments for the failure of Born's rule \cite{cmwvw,insuff,brd,ba}.

There are many logically possible replacements of Born's rule
\cite{cmwvw,insuff,brd,ba}.  Here I shall describe only three of them.  It is
simplest to start with definitions of unnormalized relative probabilities
(nonnegative, but not necessarily summing to unity) and then to define the
normalized first-person probabilities $P_j = P(O_j)$ to be these relative
probabilities divided by their sum over all possible observations.

One choice (theory $T_3$ in \cite{brd}) would be to take the relative
probabilities to be the expectation values of the numbers of occurrence of the
observation.  This rule was called {\it volume weighting}.

A second choice (theory $T_4$ in \cite{brd}) would be for the relative
probabilities to be the expectation values of the fraction of the number of
locations in which the observation occurs.  This rule was called {\it volume
averaging}.  It differs from volume weighting when the quantum state is a
superposition of different numbers of locations (e.g., different sizes for the
universe).  Volume weighting gives more weight to components of the quantum
state in which there are more locations for observers.  On the other hand, for
volume averaging, it does not matter how many locations there are in a quantum
component, but only the fraction of the number of locations where it occurs (for
fixed quantum amplitudes for the components).

A third choice (theory $T_5$ in \cite{brd}) for each relative probability would
be the expectation value of the fraction of all observations that are the one in
question.  This rule was called {\it observational averaging}.  It would seem to
be the most natural rule to use if one assumed wavefunction collapse
\cite{mw1,mw2} and multiplied the quantum probability for a particular quantum
component with the probability of randomly choosing a particular observation out
of all the observations in that quantum component.

All three of these rules may be interpreted as replacing Born's rule of the
probabilities as expectation values of projection operators $\mathbf{P}_j$ with
rules for giving the probabilities as expectation values of other operators
$\mathbf{Q}_j$, which might be called {\it observation operators}.  That is,
these rules are linear in the quantum state (an entity that assigns expectation
values to operators).  It is logically possible that the rules for extracting
observational probabilities from the quantum state are instead nonlinear
\cite{brd}, though the examples proposed so far for this seem rather contrived
and more complicated than the linear rules.  It would perhaps be most
conservative first to explore the various linear rules.

The ambiguity of what to use to replace Born's rule is reflected in the measure
problem in cosmology \cite{BLL94,Vil95a,cmwvw}.  The usual focus of the measure
problem is how to regulate the infinities that occur when one has an infinite
universe.  However, the need to replace Born's rule shows that there is an
ambiguity even for finite but large universes.  It would not be enough to have
the actual quantum state of the universe; one would also need the rules for
extracting the first-person probabilities of observations from the quantum
state.  Here I have shown that Born's rule is insufficient for getting
reasonable first-person probabilities in a universe large enough for many
identical copies of the observer.

I am grateful for discussions with Andreas Albrecht, Tom Banks, Raphael Bousso,
Sean Carroll, Brandon Carter, Ben Freivogel, Alan Guth, Daniel Harlow, James
Hartle, Thomas Hertog, Gary Horowitz, Matthew Kleban, Andrei Linde, Seth Lloyd,
Juan Maldacena, Donald Marolf, Mahdiyar Noorbala, Daniel Phillips, Joe
Polchinski, Stephen Shenker, Eva Silverstein, Mark Srednicki, Leonard Susskind,
Herman Verlinde, Alex Vilenkin, Alexander Westphal, and others.  This research
was supported in part by the Natural Sciences and Engineering Research Council
of Canada.

\end{document}